\documentclass[pre,twocolumn,showpacs,showkeys,superscriptaddress]{revtex4}
\usepackage[ansinew]{inputenc}
\usepackage{amsmath}
\usepackage{graphicx}
\usepackage{hyperref}
\usepackage{subfigure}
\usepackage{dcolumn}
\begin{document}
\newcommand{\phantomFrac}{\vphantom{\frac{}{}}}
\newcommand{\e}{\varepsilon}
\newcommand{\subst}{\left.\phantomFrac\right|}
\newcommand{\imag}{\Im m\;}
\newcommand{\real}{\Re e\;}
\newcommand{\order}{\mathcal{O}}
\newcommand{\Az}[1][n]{A^{(#1)}}
\newcommand{\Jz}[2][n]{J_{#2}^{(#1)}}
\newcommand{\Atr}[1][n]{A_\text{Tr}^{(#1)}}
\newcommand{\An}[1][n]{{A_\text{N}}^{(#1)}}
\newcommand{\Ai}[1][n]{{A_\text{I}}^{(#1)}}
\newcommand{\Ae}[1][n]{\varepsilon_{A}}
\newcommand{\Aa}[1][n]{\chi^{(#1)}}
\newcommand{\Fi}[1][n]{\ensuremath{{f_\text{I}}^{(#1)}}}
\newcommand{\Fn}[1][n]{\ensuremath{{f_\text{N}}^{(#1)}}}
\title{Symmetry breaking in driven and strongly damped pendulum}
\author{Jukka Isoh\"{a}t\"{a}l\"{a}}
\email{jisohata@student.oulu.fi}
\affiliation{Department of Physical Sciences, P.O. Box 3000, University of Oulu FIN-90014, Finland}
\author{Kirill N. Alekseev}
\email{Kirill.Alekseev@oulu.fi}
\affiliation{Department of Physical Sciences, P.O. Box 3000, University of Oulu FIN-90014, Finland}
\affiliation{Theory of Nonlinear Processes Laboratory, Kirensky Institute of Physics, Krasnoyarsk 660036, Russia}
\author{Lauri T. Kurki}
\affiliation{Department of Physical Sciences, P.O. Box 3000, University of Oulu FIN-90014, Finland}
\author{Pekka Pietil\"{a}inen}
\affiliation{Department of Physical Sciences, P.O. Box 3000, University of Oulu FIN-90014, Finland}
\date{\today}
\begin{abstract}
We examine the conditions for appearance of symmetry breaking
bifurcation in damped and periodically driven pendulum in the case of
strong damping. We show that symmetry breaking, unlike other nonlinear
phenomena, can exist at high dissipation.  We prove that symmetry
breaking phases exist between phases of symmetric normal and symmetric
inverted oscillations.  We find that symmetry broken solutions occupy
a sufficiently smaller region of pendulum's parameter space in
comparison to the statements made in earlier considerations [McDonald
and Plischke, Phys. Rev. B 27 (1983) 201].  Our research on symmetry
breaking in a strongly damped pendulum is relevant to an understanding
of phenomena of dynamic symmetry breaking and rectification in a pure
ac driven semiconductor superlattices.
\end{abstract}
\keywords{Pendulum; symmetry breaking; chaos; semiconductor superlattice; Josephson junction}
\pacs{05.45.-a, 02.30.Oz, 73.21.Cd, 74.50.+r, 72.20.Ht}
\maketitle
\section{Introduction}
Periodically forced and damped pendulum
\begin{equation}
  \label{eq:ddp}
  \ddot\theta+\gamma\dot\theta+\sin\theta=f\cos\omega t
\end{equation}
is one of the most important paradigms in the modern nonlinear science
\cite{sagdeev}. In solid state physics, the pendulum model well
describes, for instance, the nonlinear dynamics of Josephson junctions
of superconducting materials \cite{likharev,kautz}.

Fairly popular RSJ model of Josephson junction \cite{mccumber,stewrt},
which is equivalent to an underdamped pendulum, $\gamma\ll 1$,
demonstrates very rich nonlinear dynamics. Rotational states of the
pendulum, $\langle\dot\theta\rangle\not=0$ (averaging over a period
$T=2\pi/\omega$), correspond to a generation of dc voltage across the
junction driven by ac current without dc component. Moreover,
phase-locked states of the pendulum, $\langle\dot\theta\rangle=n/m$
($n$ and $m$ are integers), correspond to quantized values of dc
voltage in the Josephson junction model. This effect, known as the
inverse ac Josephson effect \cite{langenberg,levinsen}, has already
found an application in the design of modern standard of one Volt
\cite{kautz,hamilton}.  Chaotic vibrational and rotational motions of
underdamped pendulum, $\gamma\ll 1$ are also well-known in Josephson
junctions \cite{chaos-0,d'humieres} (for review see
\cite{kurkijarvi}).  Surprisingly, the optimum operating point for the
voltage standards in $(\omega,f)$--parameter space is located near a
region of chaos \cite{kautz}.  Therefore, knowledge of conditions for
transition to chaos is important for the optimization of zero-bias
voltage standard and related Josephson devices \cite{kautz,hamilton}.

One of the most often observable roads to chaos is the period doubling
scenario \cite{eckmann}. The pendulum (\ref{eq:ddp}) belongs to a
class of symmetric dynamical systems with the invariance under the
transformations $\theta\rightarrow -\theta$ and $t\rightarrow
t+T/2$. Therefore, symmetry-breaking bifurcation is a necessary
precursor to period doubling \cite{d'humieres,macdonald,swift}. This
bifurcation describes a sharp transition from a symmetric limit cycle
satisfying
\begin{equation}
  \label{eq:sym_def}
   \theta(t+T/2) = -\theta(t) + 2\pi k \quad k\text{\ is an integer}
\end{equation}
to symmetry broken limit cycles, for which this equality is
invalid. As easily seen from (\ref{eq:ddp}), a steady-state symmetric
solutions, $\theta(t)$, can have only odd harmonics of $\omega$ in
their Fourier expansion series, together with a zero harmonic
$\langle\theta\rangle$ that is equal to either $0$ or $\pi$. The
solutions with $\langle\theta\rangle=0$ are symmetric oscillations
around the stable position $\theta=0$, while
$\langle\theta\rangle=\pi$ corresponds to oscillations of inverted
pendulum. In contrast, for symmetry broken solutions both even and odd
harmonics are possible and $\langle\theta\rangle$ is some constant
different from $0$ or $\pi$. Typically, the symmetry broken
trajectories of underdamped pendulum occupy a very small range of
parameters $\omega f$ just near transition to chaos
\cite{d'humieres,macdonald,kerr}. Detailed analytical analysis of a
transition from symmetric oscillations to symmetry broken oscillations
near chaos border in the underdamped pendulum, $\gamma\ll 1$ and
$\omega<1$, is presented in \cite{miles}.

In spite of a widely spread belief that symmetry breaking is always
connected to a transition to chaos \cite{swift}, this bifurcation is
still possible for $\gamma\ll 1$ but $\omega>1$
\cite{takimo_tange}. In these conditions chaos is impossible and
symmetry breaking arises near transitions from normal to inverted
states of the pendulum \cite{takimo_tange}.

For a strong damping ($\gamma\gtrsim 1$), neither chaotic
\cite{levi,yang}, nor rotational phase locked states \cite{d'humieres}
can exist anymore. However, the presence of symmetry broken states of
overdamped pendulum, that obviously are not related to a transition to
chaos, has been briefly mentioned earlier in two papers devoted to
dynamics of Josephson junctions \cite{macdonald,octavio}. Moreover, it
has been reported \cite{macdonald} that these symmetry broken
trajectories occupy a quite large part of the parameter space
$(\omega, f)$. This interesting aspect of the pendulum dynamics at
strong dissipation did not get further attention so far, probably
because of two main reasons. First, for the Josephson junctions
pendulum's solutions at a strong damping are not physically
interesting.  Second, $\theta$ and $\langle\theta\rangle$ do not
correspond to any directly measurable physical variables in the
junctions.

However, it has been shown recently that another type of solid state
microstructures -- semiconductor superlattices subjected to a
high-frequency electric field -- can demonstrate very rich nonlinear
dynamics similar to dynamics of the Josephson junctions
\cite{ignatov93,dunlap,ignatov95,alekseev96,alekseev98,alekseev01,alekseev02a}.
In particular, an analog of inverse ac Josephson effect has been
predicted for the superlattices \cite{ignatov95,alekseev98}.
Moreover, within some reasonable approximations nonlinear dynamics of
an ac-driven semiconductor superlattice in the miniband transport
regime is governed by a periodically forced and damped pendulum
\cite{alekseev01,alekseev02a}.  For the superlattices with realistic
scattering constant \cite{unterrainer96, winnerl97}, an effective
damping in the corresponding pendulum model is not small:
$\gamma\gtrsim 1$. Overdamped pendulum also arises in the models of
lateral semiconductor superlattices \cite{alekseev02b}.

Importantly, in contrast to the case of Josephson junctions, a voltage
across a superlattice is proportional to both the velocity,
$\dot\theta$, and the coordinate, $\theta$, of the pendulum
(\ref{eq:ddp}) \cite{alekseev01}. Therefore, even if rotations are
impossible, $\langle\dot\theta\rangle\ne 0$, a dc voltage across the
superlattice can still be generated due to contributions of
symmetry-broken swinging oscillations with $\langle\theta\rangle\ne
l\pi$ ($l=0,1$). For a strong damping, this is the only mechanism that
can contribute to a rectification of THz signal in the semiconductor
superlattice \cite{alekseev01}.

The existence of physical situations, where the symmetry breaking in
pendulum at strong dissipation can be physically important, is the
main motivation of our present work. Combining the analytical
technique of truncated Fourier expansion \cite{pedersen} with
numerical simulations we find the conditions for symmetry breaking at
strong damping.
We describe a scenario of transition from symmetric to asymmetric
oscillations at a strong damping. We found that symmetry broken (SB)
oscillations form an intermediate stage between symmetric normal (N)
and symmetric inverted (I) oscillations, i.e.
N$\rightarrow$SB$\rightarrow$I$\rightarrow$SB$\rightarrow$N. Note that
it is different from transitions N$\rightarrow$SB$\rightarrow$N with a
large symmetry breaking phase even in overdamped case, which have been
reported in \cite{macdonald}. We observed a relatively small symmetry
breaking phase; with an increase of damping the ranges of driving
amplitude and frequency, resulting in symmetry breaking, decreases.
Moreover, symmetry breaking does not exist in the overdamped case
described by the first order differential equation. In this case only
normal and inverted oscillations survive. We presented simple formulas
providing a good approximation for bifurcation points between
different types of symmetric (N$\rightarrow$I) and asymmetric
(N$\rightarrow$SB, I$\rightarrow$SB) transitions in wide ranges of the
frequency and the strength of alternating force.

The organization of this papers is as follows.  We start with a
consideration of overdamped pendulum without inertial term
($\ddot\theta=0$). The analysis of this model appears to be the most
simple and transparent. In the subsequent section~\ref{sec:ddp}, we
present analytic and numerical results on symmetry breaking
bifurcations in the pendulum with inertial term. The final section of
the paper is devoted to a summary and a brief discussion on
applications in physics of semiconductors. The main text of our paper
contains only most important results with sufficient explanations.
Details of analysis and computing are presented in Appendixes:
Analytic analysis of bifurcations in first- and second order pendulum
equations (Appendix~\ref{app:det_anal}), numerical methods in use
(Appendix~\ref{app:numer}), proof of absence of hysteresis at symmetry
breaking transition (Appendix~\ref{app:no_hysteresis}) and role of
higher order harmonics at symmetry breaking (Appendix~\ref{app:foa}).

\section{\label{sec:fop}First order pendulum equation}
When damping is very strong we can neglect the second order derivative
in (\ref{eq:ddp}) and get the first order overdamped pendulum equation
\begin{equation}
  \label{eq:fop}
  \dot\theta+\beta\sin\theta=\eta\cos\omega t
\end{equation}
with $\beta=1/\gamma$ and $\eta=f/\gamma$.  Numerical integration of
this equation shows that only two types of stable periodic motion
exists: Normal and inverted modes of oscillations.  Solutions
corresponding to these modes are all symmetric with
$\langle\theta\rangle = 0$ and $\langle\theta\rangle = \pi$ for the
normal and inverted mode solutions, respectively.
\begin{figure*}
  \includegraphics{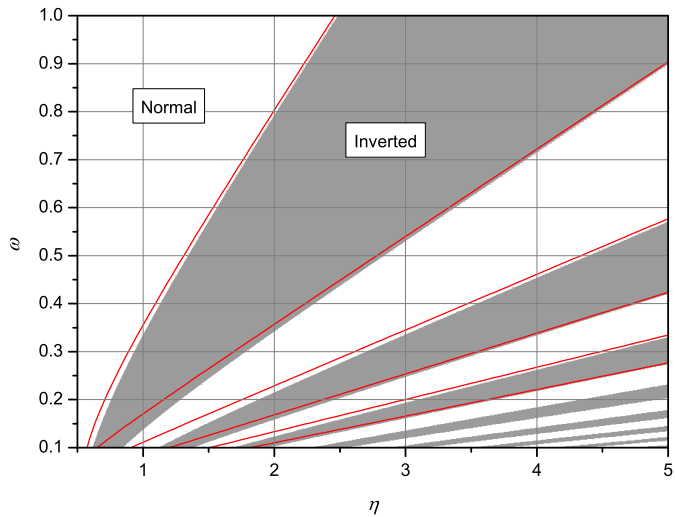}
  \caption{\label{fig:fop_2d} (Color) First order pendulum, with
    $\beta=1/2$, regions of inverted and normal mode
    oscillations. White areas correspond to normal oscillations with
    $\langle\theta\rangle=0$ and gray areas to inverted mode,
    $\langle\theta\rangle=\pi$.  No symmetry broken solutions are
    found.  Solid (red) lines indicate the analytic prediction for
    mode boundaries, Eq.~(\ref{eq:fop_boundary}), which are plotted
    for $n=1\ldots 6$.}
\end{figure*}
Figure \ref{fig:fop_2d} shows a plot of regions of the two different
modes in the $(f,\omega)$--parameter plane which exhibits a fan
shaped pattern of alternating, disjoint areas of normal and inverted
mode oscillations. A transition between the two modes is a sharp one
and no intermittent states are in spite of rigorous attempts
found. Especially, no symmetry broken regions exist.
\par
A simple analytic calculation verifies the absence of symmetry broken
solutions and provides a good approximate condition for the
transition.  The analysis is done under the assumption that the
solution is well described by the trial function
\begin{equation}
  \label{eq:fha}
  \theta=A_0 + A_1\cos(\omega t + \alpha_1).
\end{equation}
We begin by applying linear stability analysis to find the regions of
stable symmetric motion.  For details refer to
Appendix~(\ref{app:det_anal}).  Straightforward analysis leads to a
stability condition
\begin{equation}
  \label{eq:fop_stab}
  \beta\cos A_0J_0(A_1) > 0,
\end{equation}
where $J_0$ is the zeroth order Bessel function.  If for the zeroth
harmonic we have $\cos A_0<0$, then $J_0(A_1)>0$ must apply and vice
versa. Both symmetric solutions lose their stability at the same
$A_1$, which is a root of the Bessel function $J_0$.  As the Bessel
functions and the zeros of $J_0$ will appear frequently, we adopt the
shorthand notations
\begin{equation}
  \label{eq:j_sh}
  J_0(\Az[n]) = 0, \qquad \Jz{k} \equiv J_k(\Az[n]),
\end{equation}
where $\Az[n]$ is specifically the $n$th root of $J_0$ and $\Jz{k}$ is
$J_k$ evaluated at $\Az[n]$.

From the equation of motion, Eq.~(\ref{eq:fop}), we can derive using
the ansatz~(\ref{eq:fha}) relation
\begin{equation}
  \label{eq:fop_zero_alone}
  J_0(A_1)\sin A_0 = 0.
\end{equation}
From Eq.~(\ref{eq:fop_zero_alone}) it follows that if $A_0\neq 0,\pi$,
then $J_0(A_1) = 0$. However, stability condition (\ref{eq:fop_stab})
indicates that this case is not stable, as a perturbation does not
decay.  Therefore it is evident that asymmetric trajectories are not
possible.

Within the applicability of our first order approximation, the only
stable solutions are the ones with $A_0=0$ or $\pi$.  Additionally,
the pattern of normal and inverted mode solutions is revealed.
Starting from a forcing $\eta$ low enough, \emph{i.e.} for which
$J_0(A_1)>0$, Eq.~(\ref{eq:fop_stab}) implies that $A_0=0$. When $A_1$
crosses $\Az[1]$, the sign must change in both factors, leading to
$A_0=\pi$. As $\eta$ is further increased, $A_1$ reaches $\Az[2]$ and
the oscillations return to the normal mode.

The boundaries of the normal and inverted oscillations can even be
written in an explicit analytic form. From Eq.~(\ref{eq:fop}) the
expansion coefficient $A_1$ in (\ref{eq:fha}) can be found as a
function of pendulum parameters. Substituting $A_1=\Az$ we get
\begin{equation}
  \label{eq:fop_boundary}
  4\beta^2{\Jz{1}}^2 + \omega^2{\Az}^2 = \eta^2.
\end{equation}
Now we can compare our analytic results with numerical data. The
boundaries of transitions between normal and inverted modes of
oscillations, followed from Eq.~(\ref{eq:fop_boundary}), are
superimposed on the numerical results in Fig.~\ref{fig:fop_2d}. We see
that the the equality $J_0(A_1)=0$ and the prediction of
Eq.~(\ref{eq:fop_boundary}) hold well at the transition lines with
reasonable accuracy everywhere, except in the case of low
frequency. In the limiting case of very low frequency drive, the trial
solution in the simplest form (\ref{eq:fha}) can become invalid and
effects of higher harmonics should be taken into account.

\section{\label{sec:ddp}Symmetry breaking in the pendulum at strong dissipation}
\subsection{Numerical integration}
Now we return to the pendulum with inertia, Eq. (\ref{eq:ddp}).  We
start with a review of our numerical results; for a detailed
description of numerical methods see Appendix~\ref{app:numer}.  The
study of $(\omega,f)$--parameter plane reveals a fan shaped structure
of regions of normal and inverted mode oscillations, similar to that
of the first order pendulum. In addition to the alternating pattern of
the symmetric solutions, narrow regions of symmetry broken solution
have now emerged on the interfaces between the regions. For the
solutions in these regions we have observed nonzero values of zero
harmonic components along with nonzero even harmonics, which are
requisites of a symmetry-broken solution.
\begin{figure*}
  \includegraphics{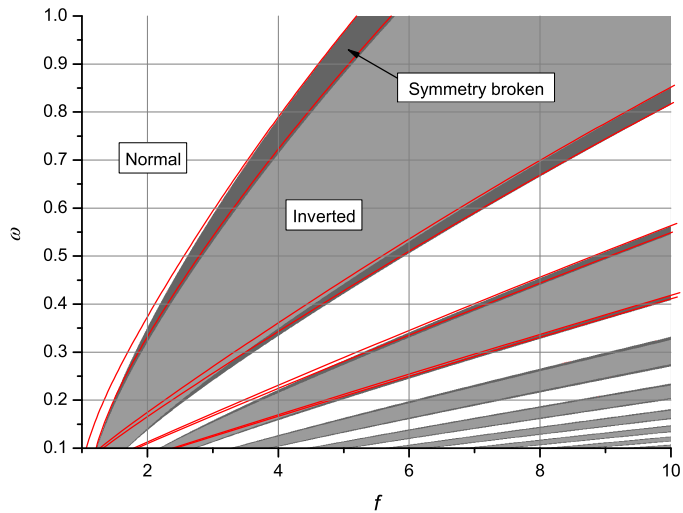}
  \caption{\label{fig:ddp_2d} (Color) Second order pendulum equation,
    regions of inverted and normal state oscillations. Sandwiched
    between them are the regions of symmetry breaking.  Solid (red)
    lines indicate analytic prediction of Eqs.~(\ref{eq:f_bif}) and
    (\ref{eq:a_bif}).  }
\end{figure*}
The different modes on the parameter plane are depicted in
Fig.~\ref{fig:ddp_2d}, where we have chosen $\gamma=2$ for the damping
coefficient. Symmetry broken regions appear around both transitions,
from normal to inverted and back, thus separating the normal and
inverted mode regions completely. Regions are not wide, in contrast to
what was reported in \cite{macdonald}
\footnote{ The pattern of symmetric inverted oscillations plus the
    nearest symmetry breaking region in our computations is found to
    be quite similar to the pattern described as ``symmetry broken''
    for overdamped pendulum in ref.\cite{macdonald}. This might be
    explained observing that technique used in \cite{macdonald} to
    detect symmetry broken trajectories at strong damping can not in
    fact distinguish symmetric inverted and instant asymmetric
    solutions.}.
Increasing the damping will cause them to get more narrow.
\begin{figure*}
  \includegraphics{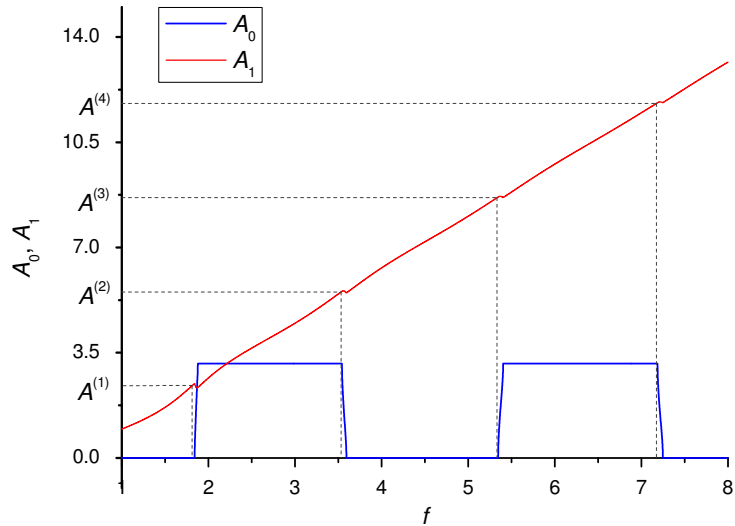}
    \caption{\label{fig:ddp_big_1d} (Color) The amplitudes of the two
    first harmonic components, $A_0$ (blue) and $A_1$ (red) plotted
    against $f$. $A_1$ shows nearly linear dependence on parameter
    $f$. Small deviations from this are seen near the transitions
    between normal and inverted states, along with a $A_0$ component
    that is not equal to $0,\pi$.  To guide the eye, dashed lines
    connecting the points $A_1(f)=\Az$ on the vertical axis to the
    corresponding $f$ on the horizontal axis are drawn.  The fixed
    pendulum parameters are $\gamma=3$ and $\omega=0.2$.}
\end{figure*}
A closer examination is shown in Fig.~\ref{fig:ddp_big_1d}, where we
have fixed $\omega=0.2$ and $\gamma=3$.  Harmonic components $A_0$ and
$A_1$ are plotted against the drive amplitude $f$, and the sequence of
modes of solutions is made more clear. Normal mode bifurcates into
asymmetric solutions when $A_1$ is close to a root of the Bessel
function $J_0(A_1)$.  This symmetry broken state persists only for a
short interval of variable $f$, ending in inverted mode, when $A_0$
has reached $\pi$.  Small deviations from approximately linear
response of $A_1$ to a change of $f$ can be seen at the transitions
from normal to inverted mode.

\begin{figure*}
  \subfigure [Low frequency and weak drive, $\gamma=3$ and $\omega=0.2$.]
  {\includegraphics{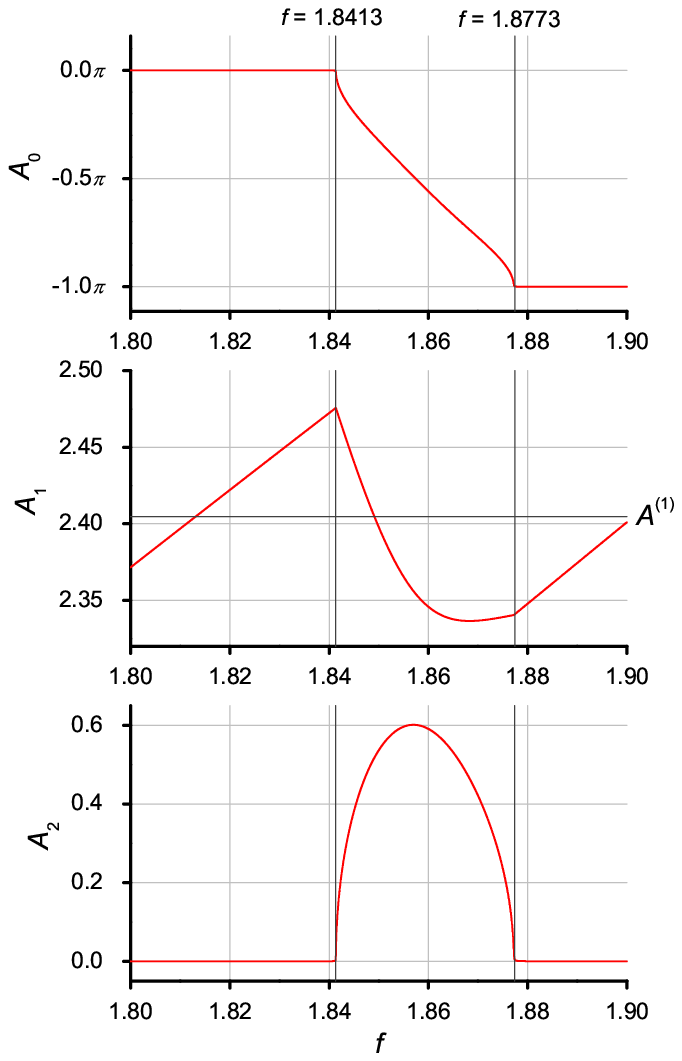}}
  \subfigure[High frequency and strong drive, $\gamma=3$ and $\omega=4$.]
  {\includegraphics{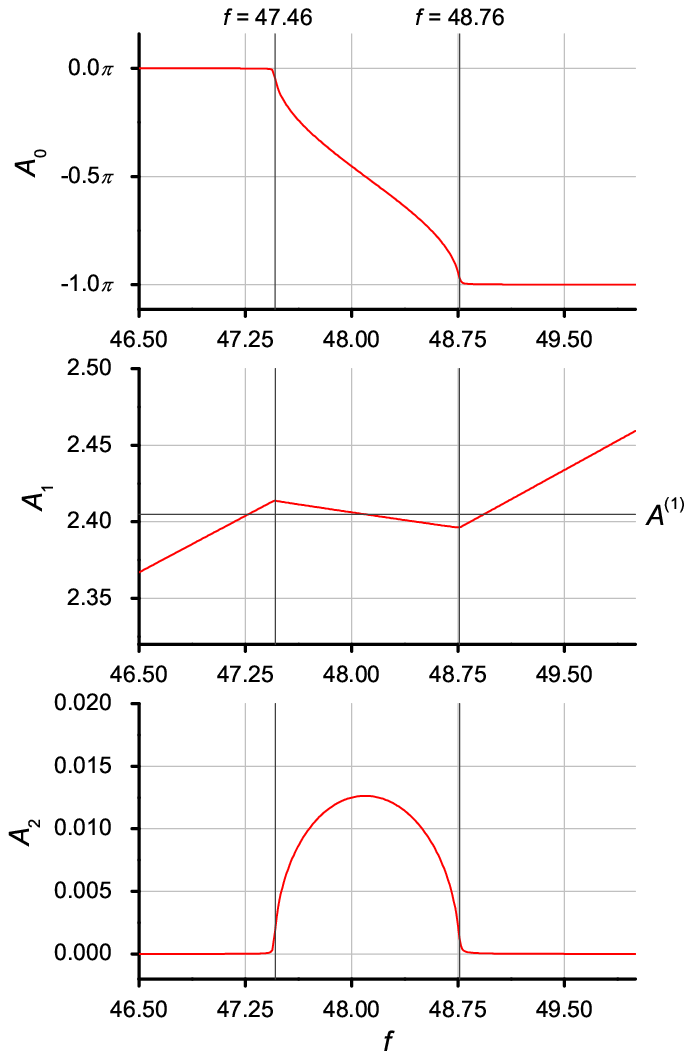}}
  \caption{\label{fig:ddp_1d_comp} (Color) The amplitudes of the two first
    harmonic components as functions of the drive amplitude $f$.
    Notice the relatively large amplitude of the second harmonic
    component in the low frequency case. }
\end{figure*}

Close-ups of the symmetry-broken range provide more insight.  The
following two examples are depicted in Fig.~\ref{fig:ddp_1d_comp}.
The first case we consider is of low frequency drive, $\omega = 0.2$
and $\gamma = 3$.
Symmetry breaking starts for moderate forcing, $f=1.8413$ and with
$A_1$ slightly larger than $\Az[1]$, the first root of $J_0$.  The
curve of $A_0$ of stable symmetric solution bifurcates from zero into
a curve corresponding to a stable asymmetric solution. This is
accompanied with the emergence of even harmonics in the Fourier
spectrum and a sudden stop in the increase of $A_1$. Further increases
of $f$ cause $A_1$ to decrease, while $A_2$ traces a sugar-loaf shaped
curve.  It should be noted, that the curve of the second harmonic
reaches quite high values, $A_2\sim 0.6$ at the maximum. The zero
harmonic varies continuously and monotonously from zero to $\pi$.  We
should note that actually two different stable solutions exist in the
symmetry breaking range, that are related to each other by the
symmetry.  Initial values of the system determine which branch the
solution converges to after the initial transient.  Symmetry breaking
ends in an inverted state at $f=1.8773$ with $A_1$ just less than
$\Az[1]$.

In the second case we examine, we set $\omega=4$ and $\gamma=3$.  To
achieve symmetry breaking, one needs high values of $f$. In this case
we have $f=47.46$. Again, $A_1$ has just crossed $\Az[1]$ when the
symmetric mode loses stability. Now the decrease of $A_1$ in the
symmetry broken region is less dramatic than in the low frequency
case. Furthermore, the second harmonic component is almost
negligible. In other respects, this case doesn't differ from the
previous example.

\subsection{\label{sec:anal_ddp}Analytic analysis}
In this subsection we present explicit analytic form of conditions for
the bifurcation to occur and provide a picture of the scenario of
symmetry breaking that is consistent with the numerical results. Here,
as in Sec.~\ref{sec:fop}, we assume that oscillations of the pendulum
can be described by the trial function in the form (\ref{eq:fha}).
For details and discussion on the range of validity of our analytic
results, see Appendix~\ref{app:det_anal}.

Following the method of \cite{pedersen,d'humieres}, we linearize
Eq.~(\ref{eq:ddp}) and, with appropriate approximations, put it into
the form of the Mathieu equation.  Using the existing knowledge of the
solutions of Mathieu equation, we derive the following conditions for
bifurcations from symmetric oscillations
\begin{equation}
  \label{eq:bif_cond}
  J_0(A_1)+(-1)^k\frac{2}{\gamma^2+4\omega^2}J_2(A_1)^2=0,
\end{equation}
where $k=0$ corresponds to a normal mode, while $k=1$ marks an
inverted mode.  Notice, that Eq.~(\ref{eq:bif_cond}) is the same
condition for loss of stability as the one for the first order
pendulum, $J_0(A_1)=0$, but now with an additional term. This term
only becomes significant when $A_1$ is close to a root of the Bessel
function $J_0$. As a consequence, now there are two separate
bifurcation points corresponding to the two different symmetric
modes. These are located rather close to the roots of $J_0$.  Clearly,
the dynamics of the second and the first order pendula are similar
away from the regions of transition, justifying the correspondence to
the analysis of the previous section.

The explicit solution of Eq.~(\ref{eq:bif_cond}) in terms of amplitude
$A_1$ can now be found.  Owing to the fact that the latter term in
Eq.~(\ref{eq:bif_cond}) is small, we get for the critical values of
$A_1$
\begin{eqnarray}
  \label{eq:a_bif}
  \left.
  \begin{array}{r} \An \\ \Ai \end{array}\right\}
  &=& \Az \pm \frac{2{\Jz{2}}^2}{\Jz{1}(\gamma^2+4\omega^2)\mp\Aa}, \\
\label{eq:chi_def}
  \Aa &=& 2{\Jz{2}}(\Jz{1}-\Jz{3}).
\end{eqnarray}
Here $\An$ and $\Ai$ are respectively the critical amplitudes of
normal and inverted modes for which symmetric oscillations become
unstable.  A small asymmetry with respect to $\Az$ in the critical
$A_1$ due to the term $\Aa$ can be neglected within the applicability
of Eq.~(\ref{eq:a_bif}) without loss of qualitative agreement. This is
implicitly assumed hereafter in this section.

In transitions from normal to the next inverted mode, critical value
of amplitude $A_1$ for the inverted mode is less than that of the
normal mode: $\Ai<\An$. The reverse applies for transitions from an
inverted phase to the next normal phase region, \emph{i.e.} $\An<\Ai$.
Therefore $A_1$ must decrease as the boundary of the first symmetry
broken region is crossed, and the $n$th region starts when
$A_1>\Az[n]$ and ends when $A_1<\Az[n]$. Such kind of behavior of
$A_1$ is in a good agreement with numerical results shown in
Fig.~\ref{fig:ddp_1d_comp}. Moreover, following Eq.~(\ref{eq:a_bif})
the difference between $\An$ and $\Ai$ decreases with an increase of
the frequency $\omega$.  This explains why the range of $A_1$ within
the symmetry breaking region is wider in the low frequency case in
comparison with the case of high frequency drive (cf subplots (a) and
(b) in Fig.~\ref{fig:ddp_1d_comp}).

We turn now to finding of surfaces in the parameter space $(\gamma,
\omega, f)$, on which the symmetry breaking bifurcation
happens. First, we substitute the trial solution (\ref{eq:fha}) in the
equation of motion (\ref{eq:ddp}) and in the first harmonic
approximation derive the dependence of amplitude $A_1$ on the pendulum
parameters $\gamma$, $\omega$ and $f$. Second, substituting critical
values of $A_1$ from Eq.~(\ref{eq:a_bif}) and taking $A_0=k\pi$, we
have
\begin{equation}
  \label{eq:f_bif}
  f^2 = (\gamma\omega A_1)^2 + \left(\omega^2A_1 \pm 2J_1(A_1)\right)^2,
  \, \left\{
  \begin{array}{ll}
  A_1 = \Ai \\
  A_1 = \An
  \end{array}
  \right.
\end{equation}
for bifurcations from inverted and normal modes, respectively.  Taking
the difference of these two critical $f$ does indicate that one
symmetric mode loses stability before the other gains it. This
eliminates a possibility of hysteresis around the symmetry breaking
transition and shows that both symmetric phases cannot be stable for
same pendulum parameters. For more detailed discussion on this
subject, see Appendix~\ref{app:no_hysteresis}.

Our main analytic results, Eqs. (\ref{eq:f_bif}) and (\ref{eq:a_bif}),
are expected to be valid for $\omega\gtrsim 1$.  Even for frequencies
less than unity, these results may be applied, provided that $f$ is
sufficiently large. Increasing the damping does somewhat improve
results, and for low frequency overdamped case, good qualitative
picture can be still provided even for low frequencies, $\omega\simeq
0.1$ (see Appendix~\ref{app:det_anal}). The effects of third and
higher order harmonics in the trial solution (\ref{eq:fha}) are
discussed in Appendix~\ref{app:foa}.

On the other hand, Eq.~(\ref{eq:f_bif}) shows remarkable agreement
with the numerical results for a very wide range of parameters. In
Fig.~\ref{fig:ddp_2d} analytic prediction for the boundaries of
symmetry breaking regions are shown as solid (red) lines.
Additionally, in Table~\ref{tab:an_comp} we have tabulated points of
symmetry breaking bifurcations in $f$ variable for a selected set of
quite different values of $\omega$ and $\gamma$. Boundaries of regions
of stable symmetric solutions can be found even in low damping and
moderately low frequency case, provided that $f$ is sufficiently
large.  This suggests that as damping or frequency are lowered from
sufficiently high values, the symmetry broken regions evolve into the
regions of complex and chaotic dynamics with the alternating normal
and inverted phase motion separating them. Evidently, our analytic
results provide a reasonable approximation even in parameter space
regimes beyond only overdamped case.

Now we can summarize the main results of present section.  Taking the
alternating force strength, $f$, as the control parameter, we can
report the following behavior: A stable inverted or normal mode
becomes unstable at an amplitude of stationary oscillations, $A_1$,
that is just greater than $\Az$ and symmetry breaking starts.
Increasing $f$ further causes $A_1$ to decrease, as the bifurcation to
the next mode should occur when $A_1$ is less than $\Az$.  The
critical values of the $A_1$ are given by Eq.~(\ref{eq:a_bif}) with
the critical parameters by Eq.~(\ref{eq:f_bif}).
These critical points approach a common value,
$\Az$, as damping is increased.

\newcommand{\fca}{f_1}
\newcommand{\fcb}{f_2}
\begin{table*}
  \begin{ruledtabular}
  \begin{tabular}{ddccdddddd}
      \multicolumn{3}{c}{Parameters}
    &
    & \multicolumn{2}{c}{Numerical}
    & \multicolumn{2}{c}{Analytic}
    & \multicolumn{2}{c}{Errors (\%)}
    \\
      \gamma
    & \omega
    & n
    &
    & \fca
    & \fcb
    & \fca
    & \fcb
    & e_\text{Max}
    & e_\text{Width}
    \\
    \hline
    2    & 1   & 1 &                  &  5.19964 &  5.76731  &  5.22416  &  5.7492  & 0.48  & 7.6  \\
    3    & 2   & 2 &                  & 39.4625  & 40.1666   & 39.4589   & 40.1622   & 0.011 & 0.11 \\
    0.05 & 2   & 1 &                  &  8.76666 & 10.51386  &  8.79088  & 10.5046   & 0.28  & 2.0  \\
    0.5  & 1   & 2 & \footnotemark[1] &  5.61680 &  6.79113  &  5.60002  &  6.76786  & 0.35   & 0.56 \\
    0.5  & 0.5 & 3 & \footnotemark[1] &  2.73651 &  3.53534  &  2.71429  &  3.45881  & 2.2   & 6.8
    \footnotetext[1]{Region between
    the critical $f$ involves in addition to symmetry broken solutions
    also complex dynamics including chaos.}
  \end{tabular}
  \end{ruledtabular}
  \caption{\label{tab:an_comp} Locations and widths of selected
  symmetry breaking intervals. Here $(\gamma, \omega, n)$ are the
  parameters of the system with $n$ being the number of the
  transition.  The start and end points of the symmetry broken range
  are, respectively, $\fca$ and $\fcb$.  In the errors' column,
  $e_\text{Max}$ is the larger of the relative errors in the start or
  end points of the symmetry breaking interval. Additionally the error
  in the width of the symmetry broken interval $e_\text{Width}$ is
  given.  Analytically given values for the start and end points are
  obtained using Eq.~(\ref{eq:f_bif}) along with Eq.~(\ref{eq:a_bif}).
  Notice the good agreement between the analytic and numerical result
  even for very low damping and moderate frequencies.}
\end{table*}

\section{\label{sec:concl}Conclusion}
In summary, we confirmed the existence of symmetry breaking phenomenon
in the strongly damped pendulum. The symmetry breaking states form a
necessary stage in transitions from the normal to the inverted and
from the inverted to the normal states of the overdamped pendulum. The
scenario of transitions from the symmetric to the symmetry broken
states, and vice versa, appears to be similar to the case of
underdamped pendulum without chaos \cite{takimo_tange}.  Our results
also demonstrate that an inertial term in pendulum is a requisite of
symmetry breaking.  On a more hypothetical note, it could then be
possible to achieve symmetry breaking similar to what it described
here, if higher even derivatives or/and additional nonlinear damping
terms are included.

These results can be directly applied to the studies of novel
mechanism of THz radiation rectification in semiconductor
superlattices \cite{alekseev98} because this system has underlying
pendulum dynamics \cite{alekseev01,alekseev02b}.  Note that in some
cases pendulum-like equation of the third order \cite{tetervov,bass86}
and a pendulum equation with a nonlinear damping term supporting
symmetry of the problem \cite{malkin} naturally arise in different
models of semiconductor superlattices.  We speculate that swinging
symmetry broken oscillations in these superlattice models also
correspond to a spontaneous generation of dc voltage or dc
current. Related detailed consideration will be published
elsewhere. It is worth to notice also the importance of the symmetry
breaking bifurcations in the physics of ac-driven bulk semiconductors
with different types of nonlinearity \cite{bumyalene}.  Our results
can be also useful in the development of this area of semiconductor
nonlinear dynamics.

\begin{acknowledgments}
This research was partially supported by the Academy of Finland, grants No 1206063 and No 100487.
\end{acknowledgments}

\appendix
\section{\label{app:det_anal}Details of analytic analysis}
\subsection{First order pendulum}
We start with a derivation of stability condition (\ref{eq:fop_stab}).
Consider a small perturbation $\delta(t)$ to a solution of
Eq.(\ref{eq:fop}). The evolution of the perturbation is governed by
the linearized equation
\begin{equation}
  \dot\delta + \beta\cos\theta\; \delta = 0.
\end{equation}
Its formal solution is
\begin{equation}
  \delta = \exp\left(-\beta\int dt\; \cos\theta\right).
\end{equation}
Using Eq.~(\ref{eq:fha}) and well known identities
\begin{equation}
  \label{eq:SinCos}
  \sin(a\cos\phi)=2\sum_{n=0}^\infty(-1)^nJ_{2n+1}(a)\cos\left((2n+1)\phi\right),
\end{equation}
\begin{equation}
  \label{eq:CosCos}
  \cos(a\cos\phi)=J_0(a)+ 2\sum_{n=1}^{\infty}(-1)^nJ_{2n}(a)\cos(2n\phi),
\end{equation}
the solution $\delta(t)$ is put into the form
\begin{equation}
  \label{eq:fop_lin_sol}
  \delta = \exp \left(-\beta\cos A_0J_0(A_1)t + \text{periodic terms}
  \right).
\end{equation}
A perturbation to a stable solution must be damped, which leads to the
stability condition (\ref{eq:fop_stab}).

Now we turn to the derivation of Eq.~(\ref{eq:fop_boundary}).  We
substitute the ansatz into Eq.~(\ref{eq:fop}).
Identities~(\ref{eq:SinCos}) and (\ref{eq:CosCos}), truncated to
include terms up to the first harmonic, are now employed to expand the
sine term. Finally, in the resulting expression equating the
trigonometric functions separately along with the zero harmonic, the
following set of equations is obtained.
\begin{subequations}
\label{eq:fop_fha}
\begin{eqnarray}
  \label{eq:fop_cos}
  -2\beta\cos A_0 J_1(A_1) &=& \eta\cos\alpha_1 \\
  \label{eq:fop_sin}
  \omega A_1 &=& \eta\sin\alpha_1 \\
  \label{eq:fop_zero}
  J_0(A_1)\sin A_0 &=& 0
\end{eqnarray}
\end{subequations}
Eliminating the phase between Eqs.~(\ref{eq:fop_cos}) and
(\ref{eq:fop_sin}) we get
\begin{equation}
\label{eq:fop_eta}
 4\beta^2\cos^2A_0J_1(A_1)^2 + \omega^2A_1^2 = \eta^2.
\end{equation}
Eq.~(\ref{eq:fop_boundary}) follows by substituting $A_1=\Az$
and $A_0=0,\pi$.

\subsection{Second order pendulum}
\subsubsection{Derivation of Eqs.~(\ref{eq:bif_cond}) and (\ref{eq:a_bif})}
Linearization of Eq.~\eqref{eq:ddp} yields the equation for the
evolution of an infinitesimal perturbation
\begin{equation}
  \label{eq:linddp}
  \ddot\delta + \gamma\dot\delta + \cos\theta\;\delta = 0.
\end{equation}
Change of variables $\delta=\exp(-\gamma t/2)\;\xi$ removes the first
order derivative term. The trial function~(\ref{eq:fha}) is again used
to substitute $\theta$ and the cosine is expanded up to second order,
as the first order term vanishes for symmetric mode
solutions. Finally, after rescaling the time $\tau=\omega t+\alpha_1$,
Eq.~(\ref{eq:linddp}) is put into the form
\begin{equation}
  \label{eq:mathieu}
  \ddot\xi + \frac{(-1)^k}{\omega^2}
  \left[
    J_0(A_1)-(-1)^k\frac{\gamma^2}{4}-2J_2(A_1)\;\cos 2\tau
  \right]\xi=0.
\end{equation}
Overdot stands for differentiation with respect to $\tau$ and we use
$A_0=k\pi$. It is easy to see that Eq.~(\ref{eq:mathieu}) has the form
of the Mathieu equation,
\begin{equation}
  \ddot y + (a-2q\cos 2\tau)y=0,
\end{equation}
when we make the identifications
\begin{eqnarray}
 a&=&\frac{(-1)^k4J_0(A_1)-\gamma^2}{4\omega^2}, \\
 q&=&\frac{(-1)^k}{\omega^2}J_2(A_1).
\end{eqnarray}
Following Floquet theorem, a solution of Mathieu equation can be
presented in the form $y=P(\tau)\exp(i\nu \tau)$, where $P$ is a
periodic function and $\nu$ is the Mathieu characteristic
exponent. Stability of the symmetric oscillations is now determined by
the condition
\begin{equation}
  \label{eq:stab}
  -\frac{\gamma}{2\omega}-\imag\nu < 0,
\end{equation}
which holds for stable solutions.  Therefore, a symmetric solution
loses stability at $\nu = -i\gamma/2\omega$. Real part of $\nu$ is
zero.

We now turn to finding the points of bifurcation.  We can make use of
the expression for $a$ as a function of $q$ and $\nu$
\cite{abramowitz} as
\begin{equation}
  \label{eq:mpc}
  a=\nu^2+\frac{q^2}{2(\nu^2-1)}+\cdots,
\end{equation}
which we truncate to include only the two first terms.  Substituting
the expressions for $\nu$, $a$ and $q$ into (\ref{eq:mpc}), we obtain
Eq.~(\ref{eq:bif_cond}), which is the condition for a bifurcation from
a symmetric solution.  Note that the truncation of series
(\ref{eq:mpc}) is valid provided that $q$ is small or $\nu^2$ is
large. These conditions are usually satisfied, independent of
$\omega$, because higher order terms are all proportional to
$1/\omega^2$. Large damping is then needed for small $\omega$ in order
to have the series converge fast enough for the truncation to apply.

Equation~(\ref{eq:bif_cond}) can be approximately solved. We
substitute $A_1=\Az+\e$ and make a series expansion up to the first
order in $\e$.
\begin{equation}
  \label{eq:bif_lin_cond}
  2(-1)^k{\Jz{2}}^2+\e\left[-(\gamma^2+4\omega^2)\Jz{1}+(-1)^k\Aa\right] = 0,
\end{equation}
where $\Aa$ is defined in Eq.~~(\ref{eq:chi_def}).  The validity of
this approach is due to the fact that the second term in
Eq.~(\ref{eq:bif_cond}) is small for wide range of parameters.  It is
obvious that $\gamma^2+\omega^2 \gg 1$ is sufficient. However, this
can be even slightly relaxed. Symmetry breaking is expected when $A_1$
is near a root of $J_0$.  Now, due to known properties of Bessel
functions, we may infer that if $A_1$ is sufficiently large and near a
root of $J_0$, it is also close to a root of $J_2$. Thus, the
discussed second term may be small even for low values of $\gamma$ and
$\omega$.

Resulting equation (\ref{eq:bif_lin_cond}) is trivial to solve for
$\e$. Solution yields
\begin{equation}
  \label{eq:e_bif}
  \e = \frac{2(-1)^k{\Jz{2}}^2}{\Jz{1}(\gamma^2+4\omega^2)-(-1)^k\Aa},
\end{equation}
from which Eq.~(\ref{eq:a_bif}) immediately follows.  Note that the
$\Aa$ in the denominator results in a slight asymmetry of the
bifurcation points with respect to $\Az$. Within the applicability of
our results, $\Aa$ is small enough to be neglected, when a more
tractable form of Eq.~(\ref{eq:e_bif}) is needed.  Some of the
quantitative agreement is lost in this approximation, but qualitative
the result is the same. This is due to the fact that in the derivation
of Eq.~(\ref{eq:bif_lin_cond}) we already assumed, that
$\gamma^2+\omega^2$ is sufficiently large, or that $J_2(A_1)$ is
small.

\subsubsection{Derivation of Eq.~(\ref{eq:f_bif})}
Substituting the trial solution $\theta=A_0+A_1\cos(\omega
t+\alpha_1)$ into the Eq.~(\ref{eq:ddp}), expanding the $\sin\theta$
term using the second order truncations of identities
(\ref{eq:CosCos}) and (\ref{eq:SinCos}) and finally equating the
harmonic terms, the following set of equations is obtained
\begin{subequations}
\label{eq:ddp_fha}
\begin{eqnarray}
  \label{eq:addp_1}
  \sin A_0\; J_0(A_1)&=&0, \\
  \label{eq:addp_2}
  2\cos A_0\; J_1(A_1) - \omega^2A_1 &=& f\cos\alpha_1, \\
  \label{eq:addp_3}
  -\gamma\omega A_1 &=& f\sin\alpha_1.
\end{eqnarray}
\end{subequations}
Eliminating phase between Eqs.~(\ref{eq:addp_2}) and (\ref{eq:addp_3})
we get
\begin{equation}
  \label{eq:fsqr_2}
  f^2 = (\gamma\omega A_1)^2 + (\omega^2 A_1 - 2 \cos A_0\;J_1(A_1))^2.
\end{equation}
Replacing $A_0$ and $A_1$ with values corresponding to the
bifurcations of normal and inverted modes, we end up with an explicit
expression for critical points, Eq.~(\ref{eq:f_bif}).

\section{\label{app:numer}On numerical methods}
In all numerical calculations, standard double precision floating
point arithmetic has been used. For the numerical integration of
system~(\ref{eq:ddp}), the equation is put into the form
\begin{equation}
\label{eq:ddp_2}
  \begin{array}{rcl}
  \dot\theta_1 &=& \theta_2, \\
  \dot\theta_2 &=& -\gamma\theta_2 - \sin\theta_1 + f\cos\omega t,
  \end{array}
\end{equation}
where $\theta_1\equiv\theta$.  Initial values $\theta_k(0)=0$,
$k=1,2$, were used in all computations.  The 4/5th order embedded
Runge-Kutta routine with Cash-Karp coefficients \cite{cash90} is used
in majority of computations in this paper. For the comparison of
analytic and numerical results on the symmetry breaking regions,
Table~\ref{tab:an_comp}, 7/8th order Runge-Kutta with Prince-Dormand
coefficients \cite{dormand81} was employed.  In variable step-size
control, a relative accuracy of around $10^{-10}$ was typically used.

Harmonic components were extracted from the numerical solution by
introducing additional differential equations of the form
\begin{equation}
  \dot \xi_0 = \theta, \qquad
  \dot \xi_k = \theta\cos k\omega t, \qquad
  \dot \zeta_k = \theta\sin k\omega t,
\end{equation}
for $k>0$. Same error bounds and step functions were used as for the
variables $\theta_1$ and $\theta_2$. Coefficients in the Fourier
series expansion
\begin{equation}
  \theta = a_0 + \sum_{k=1}^\infty \left( a_k \cos k\omega t + b_k \sin k\omega t \right)
\end{equation}
have been evaluated using
\begin{eqnarray}
  a_0 &=& \frac{1}{T} \left(\xi_0(t_0+T)-\xi_0(t_0)\right), \\
  a_k &=& \frac{2}{T} \left(\xi_k(t_0+T)-\xi_k(t_0)\right), \\
  b_k &=& \frac{2}{T} \left(\zeta_k(t_0+T)-\zeta_k(t_0)\right).
\end{eqnarray}
This simply amounts to calculating the projection of the solution onto
orthogonal base of trigonometric functions. Finally, cosine harmonic
coefficients of the series
\begin{equation}
  \theta = A_0 + \sum_{k=1}^\infty A_k \cos (k\omega t+\alpha_k),
\end{equation}
are computed using
\begin{eqnarray}
  A_k &=& \sqrt{a_k^2+b_k^2} \\
  -\alpha_k &=&
  \left\{
  \begin{array}{ll}
  \arctan\left(\frac{b_k}{a_k}\right), & a_k > 0 \\
  \arctan\left(\frac{b_k}{a_k}\right) - \pi, & a_k < 0 \\
  \frac{b_k}{|b_k|}\frac{\pi}{2}, & a_k = 0
  \end{array}
  \right.
  .
\end{eqnarray}
Initial transient phase was eliminated by first performing a period
search on the solution. A simple integration in steps of one drive
cycle, $T=2\pi/\omega$, was made and after each step the current values
of $\theta$ and $\dot\theta$ were recorded and compared against the
values of previous steps.  A period of $nT$ is assumed, if
\begin{eqnarray*}
  |\theta_k(t) - \theta_k(t-nT)| &<& \epsilon\;(|\theta_k(t)| + |\theta_k(t-nT)|), \\
\end{eqnarray*}
applies for $k=1,2$ (or just $k=1$ for the first order pendulum).
Tolerance is determined by the parameter $\epsilon$.  Typically, a
value of $\epsilon=10^{-9}$ was used.  When near bifurcation points,
lower values were needed, $\epsilon=10^{-11}$ with correspondingly
higher integrator accuracy requirements.

To determine numerically the bifurcation points reported in
Table~\ref{tab:an_comp}, a finite time maximal Lyapunov exponent was
used \cite{benettin80, shimada79}.  Because one requires possibly a
very long integration time to reach the stable limit cycle in the
vicinity of bifurcation points, it was not deemed feasible to find the
actual values of the harmonic components.  Rather, Lyapunov exponent
as a measure of the rate of convergence of a small perturbation is
good indicator of bifurcation points, as it tends to zero as the
critical point is approached.
\section{\label{app:no_hysteresis}Absence of hysteresis}
In Section \ref{sec:anal_ddp} it was stated that there is no
hysteresis around symmetry breaking transition, because one symmetric
mode loses stability before another gains it. However, actually
situation is more complicated because regions of stability of
symmetric normal and inverted modes do overlap, if one considers just
$A_1$ as the variable. In this Appendix we will show that there is no
contradiction between overlapping of regions of stability of the
symmetric modes in $(A_0,A_1)$-plane and absence of such overlapping
in $(f,\omega)$-plane.

To begin with we want to demonstrate that the stability regions in
units $A_1$ really can overlap.  If the asymmetry of the bifurcation
points with respect to $\Az$ in Eq.~(\ref{eq:e_bif}) is neglected, the
critical values of $A_1$ for stability are given by
\begin{equation}
  \label{eq:a_bif_3}
  \left.
  \begin{array}{r} \An \\ \Ai \end{array}\right\}
  = \Az \pm \Ae
\end{equation}
\begin{equation}
  \Ae = \frac{2}{\gamma^2+4\omega^2}\frac{{\Jz{2}}^2}{\Jz{1}}.
\end{equation}
Clearly, $\Ae<0$ for even and $\Ae>0$ for odd $n$. As odd $n$
correspond to transitions from normal to inverted mode
(N$\rightarrow$I) and the other way around (I$\rightarrow$N) for even
$n$, the stability regions overlap.

Now we consider the difference between critical points of stability in
variable $f$. We substitute (\ref{eq:a_bif_3}) in (\ref{eq:fsqr_2})
and expand in $\Ae$ up to first order, then substitute $k$ and $A_1$
with values corresponding to inverted and normal modes. Taking the
difference of these values gives \newcommand{\FD}{\Delta f^2}
\begin{eqnarray}
 \FD &\equiv&  {\Fi}^2 - {\Fn}^2 \\* \notag
     &\approx& -4\Ae\Az\omega^2(\gamma^2+\omega^2) \\*
     &&        + 8\Jz{1}\omega^2\Az + 8\Jz{1}\Jz{2}\Ae.
\end{eqnarray}
We next drop the last term, as it is small under the assumptions we
have made earlier in derivation of Eq.~(\ref{eq:a_bif_3}).
Overlapping can occur when $\FD$ is negative for $n=1,3,\ldots$ and
when $\FD$ is positive for $n=2,4,\ldots$. For $n$ being odd this
condition results in
\begin{equation}
  \frac{\gamma^2+\omega^2}{\gamma^2+4\omega^2}{\Jz{2}}^2
  > {\Jz{1}}^2,
\end{equation}
which holds if regions of stability overlap.  Estimating left hand
side upwards, $(\gamma^2+\omega^2)/(\gamma^2+4\omega^2) < 1$, we have
${\Jz{2}}^2 > {\Jz{1}}^2$, which is false. Analogously the case of
even values of $n$ can be considered. These considerations confirm
that both symmetric modes cannot be stable simultaneously for same
values of parameter $f$.

\section{\label{app:foa}Applicability of the first harmonic trial function}
The analytic results derived herein all rely upon the assumption that
$\theta = A_0+A_1\cos(\omega t+\alpha_1)$ is a good approximation of a
correct $T$--periodic solution or at least captures the essential
features of it. Numerical integration of systems~(\ref{eq:ddp}) and
(\ref{eq:fop}) indicate that at least when $\omega \gtrsim 1$ applies
the higher harmonic components are insignificant.  For analytic
analysis we included the third harmonic term into the trial function,
$\theta=k\pi+A_1\cos(\omega t+\alpha_1)+A_3\cos(3\omega
t+\alpha_3)$. Substituting it in motion equation (\ref{eq:ddp}) we get
\begin{widetext}
\begin{equation}
  \label{eq:a3}
  A_3 = \frac{6(-1)^k\omega^2J_0(A_1)J_1(A_3) +(-1)^l 2\omega\sqrt{-\gamma^2J_0(A_1)^2J_1(A_3)^2+(\gamma^2+9\omega^2)J_0(A_3)^2J_3(A_1)^2}}
             {3\omega^2(\gamma^2+9\omega^2)},
\end{equation}
\end{widetext}
where $l$ is an integer determined by conditions not essential to
current analysis.  Eq.~(\ref{eq:a3}) vanishes as $1/\omega^2$ for high
frequencies, in support of the numerical result. It is noteworthy that
$A_1$, and consequently $f$, are coupled to $A_3$ only via the Bessel
functions in the numerator.  Thus, high values of $f$ and $A_1$ cannot
increase significantly the third harmonic. Strong external drive
therefore makes the relative contribution of the higher harmonics
smaller.  Clearly, problems can arise when the frequency is very
low. Numerical integration also shows, that higher harmonics become
more significant and even Eq.~(\ref{eq:a3}) breaks down. Although
individual harmonic components may have relatively low values, we have
observed that the spectrum as a whole can be quite spread. Therefore,
in this limiting case an inclusion of the first two or three harmonics
into the trial function may not be enough.

Now we discuss the influence of damping.  Surprisingly, we have found
that in general an increase of damping does not always suppress the
higher harmonics. From Eq.~(\ref{eq:a3}) we have
\begin{equation}
  A_3 \propto \frac{1}{\omega\gamma}\sqrt{J_0(A_3)^2J_3(A_1)^2-J_0(A_1)^2J_1(A_3)^2}
\end{equation}
for very large damping. $A_3$ can be large even for large $\gamma$, if
frequency is low. The Bessel functions become significant and strongly
nonlinear response to an increase of $\gamma$ is expected. Numerical
integration confirms this.  Relative amplitude of the third harmonic
with respect to $A_1$ fluctuates with an increasing amplitude as
$\gamma$ is increased.  Eventually these fluctuation will die
out. There is an upper limit for $\gamma$, approximately given by
(\ref{eq:f_bif}) when $n=1$, for which a symmetry breaking bifurcation
can happen. We have observed numerically that after this final
critical $\gamma$ is crossed, fluctuation stop and the higher
harmonics slowly fade away.

In summary, the first harmonic approximation of the solution is
definitely valid when $\omega\gtrsim 1$.  Increasing damping alone
will not make the truncation more applicable in the regime, where
interesting phenomena are expected. It will improve the accuracy
of some results presented in this paper, which rely on assumption
that $\omega^2+\gamma^2$ is large. Specifically, overdamped case
is well described even for $\omega\sim 0.1$, albeit without
impressive quantitative agreement. High values of parameter $f$
result in higher relative proportion of the first harmonic
component in the solution, thus improving the approximation.


\bibliography{bibliography}

\end{document}